\documentclass[onecolumn,superscriptaddress,10pt,nofootinbib]{revtex4-2}   
\usepackage{graphicx,amssymb}
\usepackage{color}
\usepackage{bm}

\usepackage{bm}% bold math
\usepackage{amssymb,amsmath}
\usepackage{amsthm}
\usepackage {amsmath}
\usepackage{enumerate}

\usepackage{latexsym}
\usepackage{graphicx,amssymb}
\usepackage{float}
\usepackage{siunitx}
\usepackage{mathrsfs}
\usepackage{soul}
\usepackage{physics}
\usepackage{graphicx}% Include figure files
\usepackage{dcolumn}% Align table columns on decimal point
\usepackage{bm}% bold math
\usepackage[utf8]{inputenc}
\usepackage{amsfonts,amssymb,amsmath}
\usepackage[mathscr]{eucal}
\usepackage{epsfig}

\begin{document}
	\title{%Quantum non-thermalization in a bath of spins
		%\mar{Long-time relaxation in the dissipative spin-boson model with a finite number of bath oscillators}
		Long-time relaxation of a finite spin bath linearly coupled to a qubit}

	\author{Jukka P. Pekola}
	\affiliation{Pico group, QTF Centre of Excellence, Department of Applied Physics,
		Aalto University, P.O. Box 15100, FI-00076 Aalto, Finland}
	
	\author{Bayan Karimi}
	\affiliation{Pico group, QTF Centre of Excellence, Department of Applied Physics,
		Aalto University, P.O. Box 15100, FI-00076 Aalto, Finland}
	\affiliation{Helteq group, QTF Centre of Excellence,  
		Department of Physics, University of Helsinki, P.O. Box 43, FI-00014 Helsinki, Finland}
	
	\author{Marco Cattaneo}
	\affiliation{Pico group, QTF Centre of Excellence, Department of Applied Physics,
		Aalto University, P.O. Box 15100, FI-00076 Aalto, Finland}
	
	\affiliation{Algorithmiq Ltd, Kanavakatu 3C 00160 Helsinki, Finland}
	\author{Sabrina Maniscalco}
	
	\affiliation{Pico group, QTF Centre of Excellence, Department of Applied Physics,
		Aalto University, P.O. Box 15100, FI-00076 Aalto, Finland}
	
	\affiliation{Helteq group, QTF Centre of Excellence,  
		Department of Physics, University of Helsinki, P.O. Box 43, FI-00014 Helsinki, Finland}
	
	\affiliation{Algorithmiq Ltd, Kanavakatu 3C 00160 Helsinki, Finland}

	\date{\today}
	
	\begin{abstract}
		%We discuss thermalization of quantum two-level systems with the coupling Hamiltonian in the rotating wave approximation. The two main observations are as follows. First, as one would expect, a central spin (qubit) coupled to a bath of thermally populated spins relaxes exponentially towards its thermal occupation at a well characterized rate. Second, the bath spins initially in the ground state, even when mutually coupled, do not relax towards a thermal distribution, but rather form a Lorentzian distribution peaking at those spins resonant with the initially excited spin. We argue that the latter result is a consequence of lack of ergodicity in the model with population-exchange (rotating-wave) coupling Hamiltonian.
		
		We discuss the long-time relaxation of a qubit linearly coupled to a finite bath of $N$ spins (two-level systems, TLSs), with the interaction Hamiltonian in rotating wave approximation.  We focus on the regime $N\gg 1$, assuming that the qubit-bath coupling is weak, that the range of spin frequencies is sufficiently broad, and that all the spins are initialized in the ground state.  Despite the model being perfectly integrable, we make two interesting observations about the effective system relaxation. First, as one would expect, the qubit relaxes exponentially towards its zero-temperature state at a well characterized rate. Second, the bath spins, even when mutually coupled, do not relax towards a thermal distribution, but rather form a Lorentzian distribution peaked at the frequency of the initially excited qubit. This behavior is captured by an analytical approximation that makes use of the property $N\gg 1$ to treat the TLS frequencies as a continuum and is confirmed by our numerical simulations. %We argue that the latter result is a consequence of lack of ergodicity in the model with population-exchange (rotating-wave) coupling Hamiltonian.}
	\end{abstract}

	\date{\today}
	\maketitle
	%In this paper we aim to understand thermalization in a concrete closed quantum system composed of a collection of coupled qubits (quantum two-level systems). Some of the expressions apply also for other types of baths, like those formed of harmonic oscillators. We first show that if one of the TLSs is coupled to the rest of the Gibbs-distributed TLSs, that nonequilibrium qubit reaches the population determined by the temperature of the bath TLSs. This is done assuming a tunnel-like coupling Hamiltonian and the lowest order perturbation theory. Next we consider the evolution of the distribution of the whole TLS system when initially in an arbitrary state. It appears that in the same approximation as above, the energy  distribution of the oscillators does not develop towards equilibrium since each of the TLSs interacts essentially only with those TLSs having the same energy. Next we demonstrate this explicitly and numerically without any approximation for the case when only one of the TLSs is initially in the excited state, the rest of them being in the ground state. We derive an analytic approximation that captures the numerical result, demonstrating a Lorentzian distribution around the energy of the initially excited TLS, and approaching a delta distribution in the extreme weak coupling limit.
	
	\section{Introduction}
	\label{sec:intro}
	In this paper we consider a simple and paradigmatic model of an open quantum system \cite{breuer2002theory,rivas2012open}, i.e., an individual qubit linearly coupled to a bath of spins (two-level systems, TLSs) in rotating wave approximation. Contrary to the standard descriptions at the basis of the theory of open quantum systems, however, we assume that the number $N$ of spins is finite. Open quantum systems coupled to finite baths have been studied in different works \cite{Vidiella2014,Decordi2018,Riera-Campeny2021,Riera-Campeny2022} with a particular attention to their thermodynamic properties, for which the back-reaction of the system on the environment must be taken into account \cite{Esposito2003}. In the present work our aim is to investigate the long-time behavior of the populations of both the qubit and the bath spins, looking for signatures of ``relaxation'' or ``thermalization'' \cite{Gogolin2016,dalessio2016,deutsch2018} in the physical model. Since the number of bath modes is finite, the dynamics of the wave function of the total system will oscillate in a coherent way at any time $t$ and the Poincaré recurrence time $t^*$, i.e. the time at which the state of the system is identical to its initial state at time $t=0$, will be finite. However, in the regime $N\gg 1$, we observe that ``effective relaxation'' can be observed, i.e., the dynamics of the system populations almost stabilizes at a fixed value for long times.
	
\begin{figure}
		\centering
		\includegraphics [width=0.65\columnwidth] {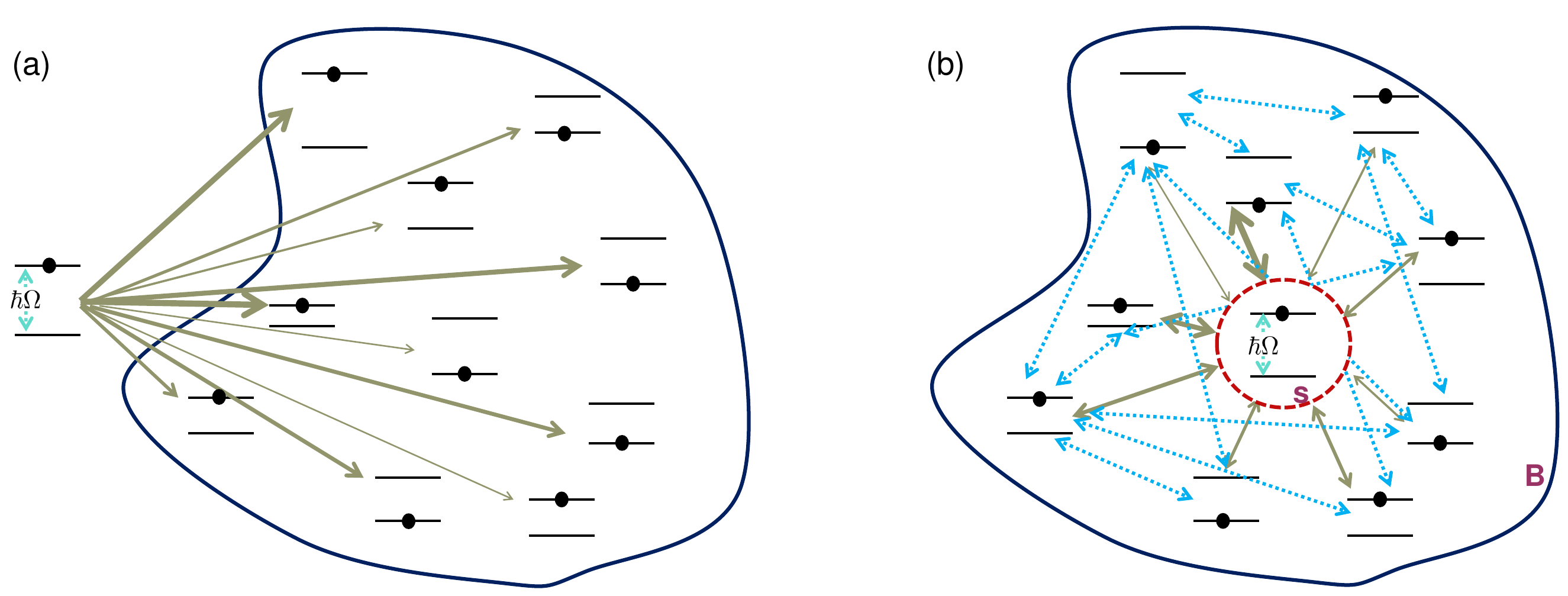}
		\caption{(a) The standard framework of an open quantum system: a qubit is coupled to a large collection of bath modes (spins, two-level systems). (b) Same as (a), but now the qubit is treated on the same footing as the bath spins. We can include inter-spin couplings in the model.  \label{Fig-sys-bath}}
	\end{figure}

	Contrary to many descriptions of equilibration and/or thermalization in closed quantum systems \cite{Gogolin2016,dalessio2016,deutsch2018,Mori_2018,Nandkishore2015,Reimann2016,Chen2021}, in this paper we do not analyse the reduced density matrix of a subsystem and we do not perform a statistical or time average. We just focus on the unitary population dynamics of the finite system depicted in Fig. \ref{Fig-sys-bath}. Our purpose is to study the problem from a different perspective, showing how even with a finite number of bath spins the qubit is for all practical purposes thermalizing, while the bath populations are relaxing towards a non-thermal distribution. We note that our analysis can be straightforwardly adapted to the spin-boson model of a qubit coupled to harmonic oscillators in a single-excitation regime.
	
	Although mainly of conceptual interest, our work can be interesting for present day solid-state qubit research. For instance, in a recent experiment the authors realised that their superconducting fluxonium qubit is coupled to an ensemble of TLSs, which in turn are fairly isolated from the true thermal bath \cite{Pop2022}. Our model could be applied to the analysis of their system, either directly or possibly with adaptation due to the weak TLS coupling to the actual bath. 
	
We dedicate this article to the memory of G\"oran Lindblad, an outstanding scientist who influenced the work of so many, and whose tools we use practically every day.
	
	We structure the rest of the paper as follows. In Sec.~\ref{sec:model} we describe the model and the Hamiltonian that we work with. In Sec.~\ref{sec:Ther-qubit} we briefly recall the standard description of the dissipative model with an infinite number of bath modes and under the standard Markovianity assumption. Then, we present our main results in Sec.~\ref{sec:singleEx}. Finally, we discuss the results and draw some conclusions in Sec.~\ref{sec:conclusions}.

	\section{Physical model}
	\label{sec:model}
	Let us now formalize the physical model. We consider a qubit with energy gap $\hbar\Omega$ coupled to a (finite) bath formed of $N$ spins with level spacing of the $j$th one equal to $\hbar\omega_j$ \cite{Pekola2022}. %\mar{While we will focus on harmonic oscillators only, the same analysis can be straightforwardly performed for spin baths as well.}
	The Hamiltonian reads $H = H_Q + V + H_B$.
	Here $H_Q=\hbar\Omega a^\dagger a$ is the qubit Hamiltonian, where $a=|g\rangle\langle e|$, with $|g\rangle,|e\rangle$ the ground and excited state, respectively. $H_B = \sum_{j=1}^N \hbar\omega_j b_j^\dagger b_j + H_B^{(int)}$ is the Hamiltonian of the bath ($H_B^{(int)}$ is a term that describes internal interactions between the bath spins, which we will specify later), where $b_j$ are annihilation operators for the spins,
	%\footnote{\mar{Note that the same notation can be straightforwardly employed to describe a spin bath, where instead of bosonic operators we have $b_j=\ket{g}_j\!\bra{e}$ for the $j$th bath spin.}}
	while $V$ is the coupling between the qubit and the bath. 
	
	Since a physical coupling, for instance inductive  or capacitive coupling in circuit QED \cite{Wendin2007a,Blais2020}, would yield $V=\sum_{j=1}^N \gamma_j (a+a^\dagger)(b_j+b_j^\dagger)$ with coupling constant $\gamma_k$ (e.g., the ``mutual inductance'' of the inductive coupling), we may take, ignoring fast rotating terms straight away,
	\begin{equation} \label{e2}
		V=\sum_{j=1}^N \gamma_j (a b_j^\dagger + a^\dagger b_j).
	\end{equation}
	We have assumed real $\gamma_k$ for simplicity.
	This ``tunneling'' coupling reads
	\begin{equation} \label{e4}
		V_I(t)=\sum_{j=1}^N \gamma_j (a b_j^\dagger e^{-i(\Omega - \omega_j)t} + a^\dagger b_j e^{i(\Omega - \omega_j)t})
	\end{equation}
	in the interaction picture with respect to the non-interacting system. %We first adopt this form of coupling in our calculations, but later extend it to include inter-oscillator couplings of the same type in Section \ref{Section3}, and comment on the limitations of these number-conserving coupling Hamiltonians in the discussion section.
	
	We may also want to include internal couplings between the bath spins. In this case, we will employ the following Hamiltonian:
	\begin{equation}
		H_B^{(int)} = \sum_{j\neq k}^{N}\kappa_{jk}\hat{b}_j^\dagger\hat{b}_k,
	\end{equation}
	which is still quadratic and in the rotating wave approximation. For simplicity, the inter-oscillator couplings $\kappa_{jk}$ are real.
	
	\section{Qubit thermalization in a Markovian scenario}\label{sec:Ther-qubit}
	In this section we recall how a single qubit coupled to an infinite thermal reservoir relaxes towards equilibrium, following the standard derivations of an open quantum system. We proceed in the weak coupling and Markovian regime for this system. For simplicity, we make the internal couplings of the bath to vanish, $H_B^{(int)}=0$. 
	%However, in this order they would not influence the final results. 
	The master equation for the density operator $\rho(t)$ of the full system and bath in the interaction picture is obtained formally from
	\begin{equation} \label{m15a}
		\dot\rho(t) =-\frac{i}{\hbar}[V_I(t),\rho(0)]-\frac{1}{\hbar^2}\int_{0}^t dt'  \big [ V_I(t),[V_I(t'),\rho(t')]\big ].
	\end{equation}
	We will assume that the initial state of the qubit and reservoir are separable: $\rho(0)=\rho^{(Q)}(0)\otimes \rho_B$, where $\rho_B$ is the thermal state of the bath. Under this assumption we observe ${\rm Tr}_B ([  V_I(t),\rho(0)])=0$.
	The reduced density operator $\rho^{(Q)}(t)$ is obtained by tracing over the spin bath degrees of freedom as $\rho^{(Q)}(t) = {\Tr}_B (\rho(t))$. 
	
	Starting from Eq.~\eqref{m15a} we can apply the standard Born-Markov approximations, whose derivation and details can be found, for instance, in Refs.~\cite{breuer2002theory,rivas2012open,lidar2019lecture}. The assumptions at the basis of such approximations are the following. i) The qubit-bath coupling constant is weak, such that the interaction Hamiltonian can be treated as a perturbation of the model Hamiltonian and we can truncate any terms in the master equation that are beyond the second order; ii) The autocorrelation functions of the bath decay sufficiently fast in time with respect to the relaxation timescale, so that the ``memory'' of the past dynamics is lost and the evolution of the state of the system is fully Markovian. The latter assumption is satisfied only if the bath is infinite, so that the recovery time of the bath oscillations is infinite.
	
	After applying the Born-Markov approximations the master equation reads
	\begin{equation} \label{m15}
		\dot\rho^{(Q)}(t) =-\frac{1}{\hbar^2}{\rm Tr}_B\Big \{\int_{-\infty}^t dt'  \big [ V_I(t),[V_I(t'),\rho^{(Q)}(t)\otimes \rho_B]\big ] \Big \}.
	\end{equation}
	Here ${\rm Tr}_B$ refers to the partial trace over the degrees of freedom of the bath. The basic master equation for the qubit is obtained from Eq. \eqref{m15}. We skip here all the details of the derivation that can be found in Refs.~\cite{breuer2002theory,rivas2012open,lidar2019lecture}. Focusing on the population of the ground state $\rho^{(Q)}_{gg}=1-\rho^{(Q)}_{ee}$, we have
	\begin{equation}
		\label{eqn:detailBalance}
		\dot\rho^{(Q)}_{gg}(t) = -\Gamma_\uparrow \rho^{(Q)}_{gg}(t) + \Gamma_\downarrow \rho^{(Q)}_{ee}(t).
	\end{equation}
	The transition rates are then given by
	\begin{equation} \label{m22}
		\Gamma_\downarrow = \frac{2\pi}{\hbar^2} \sum_{j =1}^\infty \gamma_j^2 \,  {\rm Tr}_B (\rho_B  b_{j}b_j^\dagger)\delta(\omega_j-\Omega)	,
		\quad \Gamma_\uparrow = \frac{2\pi}{\hbar^2} \sum_{j =1}^\infty \gamma_j^2 \,  {\rm Tr}_B (\rho_B b_j^\dagger b_{j})\delta(\omega_j-\Omega)
		.
	\end{equation}
	For spins in equilibrium we have
	\begin{equation}
		\Gamma_\downarrow = \frac{2\pi}{\hbar^2}\sum_{j=1}^\infty \gamma_j^2 [1-f(\hbar\omega_j)] \delta(\omega_j-\Omega), \qquad  \Gamma_\uparrow = \frac{2\pi}{\hbar^2}\sum_{j=1}^\infty \gamma_j^2 f(\hbar\omega_j) \delta(\omega_j-\Omega).
	\end{equation}
	Here $f(\hbar\omega_j)$ is the Fermi-Dirac occupation number $f(\hbar\omega_j)=1/(1+e^{\beta\hbar\omega_j})$, where $\beta=1/k_B T$ and $T$ is the temperature of the bath. For convenience, we also introduce the decay rate at zero temperature
	\begin{equation}
		\label{eqn:zeroTempdecay}
		\Gamma_0 = \frac{2\pi}{\hbar^2}\sum_{j=1}^\infty \gamma_j^2  \delta(\omega_j-\Omega).
	\end{equation}
	The Dirac deltas will disappear once we transform the summation into an integral, and they will select the proper value of the coupling constant $\gamma_j^2$, with $j$ such that $\omega_j=\Omega$.
	The detailed balance condition $\Gamma_\downarrow/\Gamma_\uparrow=e^{\beta\hbar\Omega}$ is satisfied, hence the master equation~\eqref{eqn:detailBalance} drives the qubit towards its stationary thermal state at the bath temperature $T$.
	
	In this section we have focused on the state of the qubit only, while the main aim of the present work is to study the populations of the spins of the bath. Therefore, let us finally spend a few words about the state of the bath in the standard description of the Markovian model \cite{breuer2002theory}. Under this framework, the bath is supposed to be a perfect thermal reservoir made of infinite number of modes. The state of the bath $\rho_B$ is stationary with respect to the bath Hamiltonian $H_B$. Moreover, the back-reaction of the qubit on the bath is always neglected, so that the state of the bath can be approximated as $\rho_B$ at all times, which is consistent with the assumption of thermal reservoir. As a consequence, the typical studies in open quantum systems do not focus on the state of the reservoir as a function of time at all. All this is usually justified through the fact that the qubit-bath coupling is weak and that the bath is large enough to immediately absorb and ``wash away'' all the external perturbations. However, it has already been pointed out that assuming that the state of the qubit and the state of the bath are almost uncorrelated at all times is not very rigorous, as correlations (and thus perturbations of the thermal state $\rho_B$) grow linearly with time \cite{Rivas2010a}. Therefore, the usual assumption about qubit-bath factorization at all times has to be understood as a heuristic guess we need in order to apply with the Born approximation (truncating beyond the second order) without relying on the more formal projector operator techniques \cite{Rivas2010a}. 
	
	\section{Long-time relaxation: populations of the qubit and bath spins}
	\label{sec:singleEx}
	In this section we discuss the numerically exact solution of the Schr\"odinger equation for a reservoir with a finite number $N$ of spins and compare it to an analytic approximation. Our aim is to find the long-time behavior of the qubit and bath populations in the regime $N\gg 1$.
	We will work in the interaction picture, so that we have $i\hbar \partial_t|\psi(t)\rangle=V_I(t)|\psi(t)\rangle$, where the interaction Hamiltonian is
	\begin{equation} \label{couplingall}	V_I(t)=\sum_{j=1}^{N}\gamma_j(\hat{a}^\dagger\hat{b}_j e^{i(\Omega-\omega_j)t}+\hat{a}\hat{b}_j^\dagger e^{-i(\Omega-\omega_j)t})+\sum_{j\neq k}\kappa_{jk}\hat{b}_j^\dagger\hat{b}_k e^{i(\omega_j-\omega_k)t}.
	\end{equation}
	Here $a,a^\dagger$ refer to the qubit that we will initialize in the excited state, while the fermionic operators of the bath spins are given by $b_j,b_j^\dagger$. The corresponding coupling strengths are $\gamma_j$ and $\kappa_{jk}$.
	To keep the problem tractable, we assume that all the spins are in the ground state initially, and only the qubit is excited. Because the coupling conserves the number of excitations, we may then use the basis formed of the states $\{|0\rangle=|1000...0\rangle,~|1\rangle=|0100...0\rangle,...,|j\rangle=|0~0...1^{( j\text{th})}...0\rangle,...\}$, where the first entrance refers to the qubit which is initially in the excited state $|e\rangle$ and from the second on to each of the $N$ spins in the bath.  
	
	The model we have introduced above in the single-excitation regime is exactly integrable\footnote{While the notion of integrability in quantum systems is not always well-defined \cite{Gogolin2016}, here we use it with a quite simple meaning: we can find $N+1$ commuting and ``independent'' conserved quantities of the physical model.}. We can realize this by noticing that the interaction Hamiltonian in Eq.~\eqref{couplingall} can be trivially diagonalized by finding the eigenmodes of the total system, each of which will oscillate with a finite frequency (see Appendix~\ref{appendix:integrability} for details). Then, the state of the system at time $t$ will simply be given by the sum of all the oscillatory modes with suitable weights. This means that no strict relaxation (let alone thermalization) can emerge in this model. However, we will see that a kind of ``effective'' relaxation can be observed in the regime $N\gg 1$. To do so, we next study the dynamics of the model by writing the Schr\"odinger equation for the single-excitation basis.
	
	The time evolution of the state of the whole system (qubit + $N$ spins) in the interaction picture, $|\psi(t)\rangle=\sum_{j=1}^{N}\mathcal{C}_j(t)|j\rangle$, then reads  
	\begin{eqnarray}\label{ampsint}
		&&i\hbar\dot{\mathcal{C}}_0=\sum_{j=1}^{N}\gamma_j e^{i(\Omega-\omega_j)t}\mathcal{C}_j\nonumber\\
		&&i\hbar\dot{\mathcal{C}}_j=\gamma_j e^{-i(\Omega-\omega_j)t}\mathcal{C}_0+\sum_{j\neq k}\kappa_{jk}e^{i(\omega_j-\omega_k)t}\mathcal{C}_k.
	\end{eqnarray}
	With the given initial conditions $\mathcal{C}_0(0)=1$ and $\mathcal{C}_j(0)=0$ for $j=1,\,...\,,\, N$, i.e. with state $\ket{\psi(0)}=|0\rangle$, we find 
	\begin{eqnarray}\label{cit}
		&&\mathcal{C}_0(t)=1-\frac{i}{\hbar}\sum_{j=1}^{N}\gamma_j \int_{0}^{t}dt' e^{i(\Omega-\omega_j)t'}\mathcal{C}_j(t')\nonumber\\&&\mathcal{C}_j(t)=-\frac{i}{\hbar}\gamma_j\int_{0}^{t}dt'\,e^{-i(\Omega-\omega_j)t'}\mathcal{C}_0(t')-\frac{i}{\hbar}\sum_{k\neq j}^{N}\kappa_{jk}\int_{0}^{t}dt'e^{i(\omega_j-\omega_k)t'}\mathcal{C}_k(t').
	\end{eqnarray}
	These equations can then be exactly solved numerically. As will be shown in the next section, we can also find a simple analytical result for the model populations at time $t$ in the limit of weak couplings $\gamma_j$, no internal couplings $\kappa_{jk}$, and for $N\gg 1$, as we show in the following.
	
	\begin{figure}
		\centering
		\includegraphics [width=0.9\columnwidth] {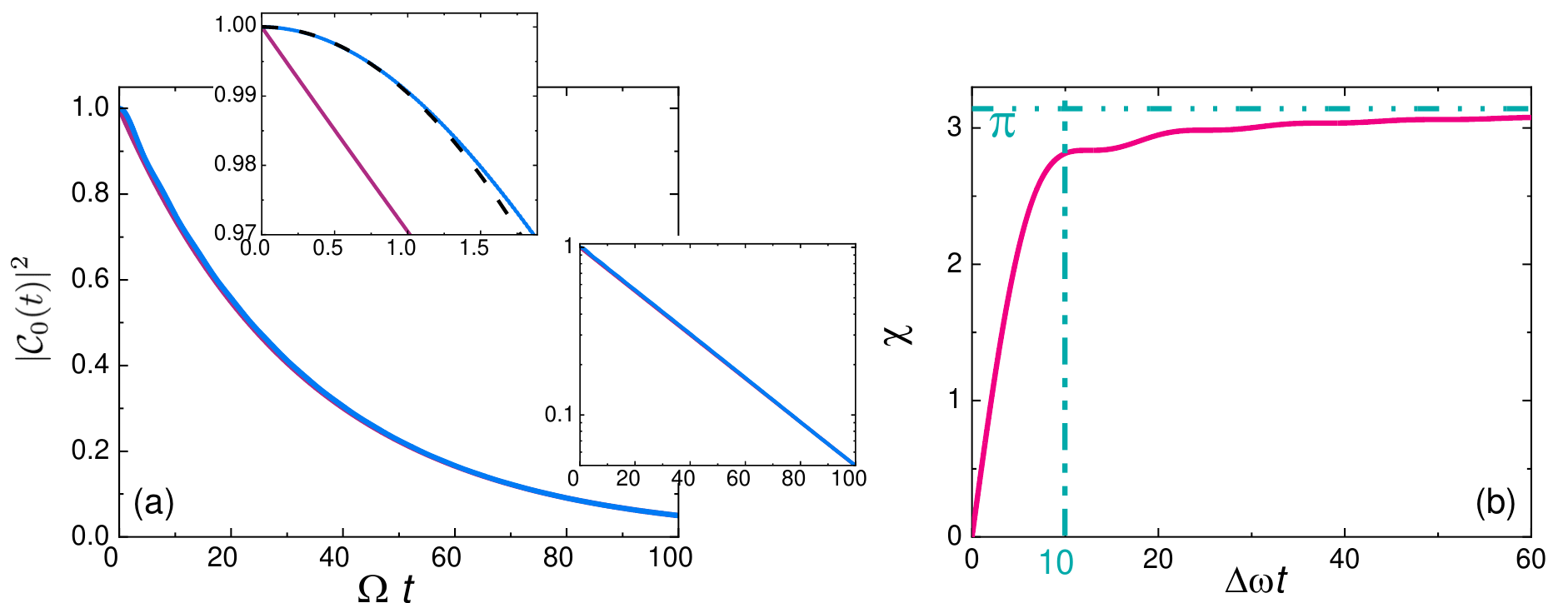}
		\caption{Relaxation of the qubit in different regimes. (a) The solid blue line shows the full solution of the Schr\"odinger equation for $N=10^6$ bath oscillators. The parameters are $\Gamma_0 =0.03\Omega$, $\Delta \omega=2\Omega$, $\kappa_{ij}= 0$. The red line shows the exponential decay, $|\mathcal{C}_0(t)|^2\simeq\exp(-\Gamma_0 t)$ closely following the numerical line at times $\Omega t\gg 1$. The upper inset presents the short time decay, together with the quadratic approximation of Eq.~\eqref{cit2_2} shown by the nearly overlapping black dashed line. The lower inset displays the same data as in the main panel on the logarithmic vertical scale. (b) The sum of Eq.~\eqref{cit2_1} for a uniform energy distribution of independent bath oscillators in form $\chi(t)=\int_{-\Delta\omega t/2}^{\Delta\omega t/2}dw (1-\cos w)/w^2$ leading to quadratic dependence [Eq.~\eqref{cit2_2}] of $|\mathcal{C}_0(t)|^2$ on $t$ at short times $\Delta\omega t\lesssim 10$, and linear and eventually exponential $t$ dependence [Eq.~\eqref{cit2_3}] at large times $\Delta\omega t\gtrsim 10$. 	\label{Fig-decay} }
	\end{figure}
	
	\subsection{Analytical approximation}
	We will for now neglect the internal coupling between the spins, that is, $\kappa_{jk}=0$. Moreover, we consider $\gamma_j$ to be a small perturbation of the model Hamiltonian, i.e., weak coupling between qubit and bath. Then, we can solve Eq.~\eqref{cit} iteratively. After inserting the initial conditions to the right-hand side of~\eqref{cit}, let us focus on the behavior of $\mathcal{C}_j(t)$. With $\mathcal{C}_0(t')\approx 1$, we find
	\begin{equation}
		\mathcal{C}_j^{(0)}(t)=\frac{\gamma_j}{\hbar}\frac{e^{-i(\Omega-\omega_j)t}-1   }{\Omega-\omega_j},
	\end{equation}
	where the superscript in parenthesis refers to the lowest iteration order.
	Using $|\mathcal{C}_0^{(0)}(t)|^2=1-\sum_{j=1}^{N}|\mathcal{C}_j^{(0)}(t)|^2$, we then obtain
	\begin{equation}\label{cit2_1}
		|\mathcal{C}_0^{(0)}(t)|^2=1-\frac{2}{\hbar^2}\sum_{i=1}^{N}\gamma_i^2\frac{1-\cos((\omega_i-\Omega)t)}{(\omega_i-\Omega)^2}.
	\end{equation}  
	
	We now perform an important step to effectively introduce dissipation in the model. Assuming $N\gg 1$ and that the frequencies of the bath spins $\omega_j$ are distributed uniformly around the qubit frequency $\Omega$, we replace the sum over the oscillators with an integral as $\sum_{j=1}^{N}\rightarrow \nu_0\int_{-\frac{\Delta\omega}{2}}^{\frac{\Delta\omega}{2}}d\omega$, where $\omega$ replaces $\omega_j-\Omega$, $\nu_0=N/\Delta\omega$ is the density of oscillators in $\omega$, and $\Delta\omega$ denotes the width of the uniform distribution of $\omega_j$ symmetrically around $\Omega$. We then have $|\mathcal{C}_0^{(0)}(t)|^2=1-\frac{2}{\hbar^2}\nu_0t \langle \gamma_i^2\rangle\int_{-\frac{\Delta\omega}{2}t}^{\frac{\Delta\omega}{2}t}dw\frac{1-\cos w}{w^2}$, where $\langle . \rangle$ refers to the average over the spins, and we have assumed that $\gamma_i$ and $\omega_i$ are uncorrelated. The integral 
	\begin{equation}
		\label{eqn:chiIntegral}
		\chi(t)=\int_{-\frac{\Delta\omega}{2}t}^{\frac{\Delta\omega}{2}t}dw\frac{1-\cos w}{w^2}
	\end{equation} 
	is shown in Fig.~\ref{Fig-decay}(b). For short times the integral gives $\chi (t)\propto t$ and beyond that $\chi (t)\simeq$~constant, yielding the two main regimes (1 and 2) of relaxation as described below. 
	
	{\bf Regime 1, $\Delta\omega t\lesssim 1$:} For short times, we have the quadratic “Zeno” result
	\begin{equation}\label{cit2_2}
		|\mathcal{C}_0^{(0)}(t)|^2=1-\Lambda_0^2~t^2,
	\end{equation} 
	where $\Lambda_0^2=N\langle \gamma_i^2\rangle/\hbar^2~$. The result of this quadratic behaviour is shown by the dashed line in the upper inset of Fig.~\ref{Fig-decay}(a); it nearly overlaps with the full numerical result shown by the blue line as in the main panel.
	
	{\bf Regime 2, $\Delta\omega t \gg 1$:} In this regime, one observes the linear dependence based on the present approximation as
	\begin{equation}\label{cit2_3}
		|\mathcal{C}_0^{(0)}(t)|^2=1-\Gamma_0 t,
	\end{equation} 
	where
	\begin{equation}\label{rate}
		\Gamma_0= \frac{2\pi}{\hbar^2}\nu_0\langle \gamma_i^2\rangle
	\end{equation} 
	is the actual decay rate of the qubit. This quantity is the equivalent of Eq.~\eqref{eqn:zeroTempdecay} for the Markovian case. The last step in Eq.~\eqref{rate} again applies for a uniform distribution of spins as given above. Equation~\eqref{cit2_3} presents the linear approximation of the exponential decay which can be obtained by standard perturbation theory as well (see, e.g. \cite{Anastopoulos}). For instance, we can write the persistence amplitude of the qubit in the Schr\"odinger picture as
	\begin{equation}\label{interest}
		\mathcal{C}^{\rm (S)}_0(t)\equiv\langle 0|e^{-i\mathcal{\hat{H}}t/\hbar}|0\rangle,
	\end{equation}
	where the evolution operator reads
	\begin{equation}\label{resolvent2}
		e^{-i\mathcal{\hat{H}}t/\hbar}=\frac{i\hbar}{2\pi} \lim_{\epsilon\rightarrow 0}\int_{-\infty}^{\infty}d\omega \frac{e^{-i\omega t}}{\hbar\omega+i\epsilon-\mathcal{\hat{H}}}.
	\end{equation}
	The integrand contains an operator which can be written as
	\begin{eqnarray}\label{series2} 
		\frac{1}{z-\mathcal{\hat{H}}}=\frac{1}{z-\mathcal{\hat{H}}_0}\sum_{k=0}^{\infty}(\hat{V}\frac{1}{z-\mathcal{\hat{H}}_0})^k
	\end{eqnarray}
	with $z=\hbar\omega+i\epsilon$. For uncoupled bath oscillators $\kappa_{ij}=0$, the matrix element of the operator in the integrand, $\langle 0|(\hbar\omega+i\epsilon-\mathcal{\hat{H}})^{-1}|0\rangle$ can be easily re-summed in all orders yielding
	\begin{equation}\label{Rtot2}
		\langle 0|\frac{1}{\hbar\omega+i\epsilon-\mathcal{\hat{H}}}|0\rangle=\frac{1}{z-\hbar\Omega-\Sigma_0(z)}, 
	\end{equation}
	with self-energy function of the state $|0\rangle$ as
	\begin{equation}\label{selfE}
		\Sigma_0(z)=\sum_{i=1}^{N}\frac{\gamma_{i}^2}{\hbar\omega+i\epsilon-\hbar\omega_i}\simeq\frac{2}{\pi}\hbar \Gamma_0 \frac{\omega-\Omega}{\Delta\omega}-i\frac{\hbar\Gamma_0}{2}.
	\end{equation} 
	Here the real part is an approximation for $|\omega-\Omega|\ll \Delta\omega$ but imaginary part is exact. Combining \eqref{resolvent2}-\eqref{selfE} in $\mathcal{C}_0(t)$ we have in the limit $\frac{\Gamma_0}{\Delta\omega}\ll 1$
	\begin{eqnarray}\label{A2tot2}
		\mathcal{C}^{\rm (S)}_0(t)\simeq e^{-i\Omega t}e^{-\frac{\Gamma_0}{2}t}.
	\end{eqnarray}
	It confirms that the qubit decays exponentially as
	\begin{eqnarray}\label{A2tot3}
		|\mathcal{C}^{\rm (S)}_0(t)|^2\simeq e^{-\Gamma_0t}.
	\end{eqnarray}
	This result is naturally the same in the interaction picture $|\mathcal{C}_0(t)|^2\simeq e^{-\Gamma_0t}$. The purple line in the main panel of Fig.~\ref{Fig-decay}(a) shows Eq.~\eqref{A2tot3} with the given parameters following closely the numerical result shown by solid blue line. For very long times, $\Delta\omega t \gg \frac{\Delta\omega}{\Gamma_0}\ln  \frac{\Delta\omega}{\Gamma_0}$, the decay becomes non-exponential. We will not discuss this regime further here.
	
	Next we approximate within the second equation of Eq. \eqref{cit} in the lowest order
	\begin{eqnarray}\label{cit-2}
		&&\mathcal{C}_0(t')\simeq e^{-\Gamma_0 t'/2}\nonumber\\&&\mathcal{C}_k(t')\simeq -\frac{i}{\hbar}\gamma_j\int_{0}^{t'}dt''\,e^{-i(\Omega-\omega_j)t''}e^{-\Gamma_0 t''/2}.
	\end{eqnarray} 
	Here we again take for illustration a uniform distribution of spin frequencies centered around the qubit frequency and with width $\Delta\omega$, such that $\nu_0=N/\Delta\omega$. Substituting Eq.~\eqref{cit-2} in the second line of Eq.~\eqref{cit} with $\kappa_{jk}=0$, we have
	\begin{equation}\label{cjt}
		\mathcal{C}_j(t)\simeq -\frac{i}{\hbar}\gamma_j\int_{0}^{t}dt'\,e^{-[\frac{\Gamma_0}{2}+i(\Omega-\omega_j)]t'}= -\frac{i}{\hbar}\gamma_j\frac{1-e^{-[\frac{\Gamma_0}{2}+i(\Omega-\omega_j)]t}}{\frac{\Gamma_0}{2}+i(\Omega-\omega_j)}.
	\end{equation} 
	%The integrals in the second term of the above equation yield
	%\begin{eqnarray}\label{integrals-cjt}
	%&&\int_{0}^{t}dt'e^{i(\omega_j-\omega_k)t'}\int_{0}^{t'}dt''\,e^{-[\frac{\Gamma_0}{2}+i(\Omega-\omega_k)]t''}=\int_{0}^{t}dt'e^{i(\omega_j-\omega_k)t'}\frac{e^{-[\frac{\Gamma_0}{2}+i(\Omega-\omega_k)]t'}-1}{-[\frac{\Gamma_0}{2}+i(\Omega-\omega_k)]}\nonumber\\&&=\frac{1}{\frac{\Gamma_0}{2}+i(\Omega-\omega_k)}\int_{0}^{t}dt'[e^{i(\omega_j-\omega_k)t'}-e^{-[\frac{\Gamma_0}{2}+i(\Omega-\omega_k)]t'}\nonumber\\&&=\frac{-i}{[\frac{\Gamma_0}{2}+i(\Omega-\omega_k)](\omega_j-\omega_k)}\,(e^{i(\omega_j-\omega_k)t}-1)+\frac{1}{[\frac{\Gamma_0}{2}+i(\Omega-\omega_k)][\frac{\Gamma_0}{2}+i(\Omega-\omega_j)]}\,(e^{-[\frac{\Gamma_0}{2}+i(\Omega-\omega_j)]t}-1)
	%\end{eqnarray} 
	%\begin{eqnarray}\label{cjt-2}
	%	\mathcal{C}_j(t)&&\simeq -\frac{i}{\hbar}\gamma_j\frac{1-e^{-[\frac{\Gamma_0}{2}+i(\Omega-\omega_j)]t}}{\frac{\Gamma_0}{2}+i(\Omega-\omega_j)}.%-\frac{1}{\hbar^2}\sum_{k\neq j}^{N}\kappa_{jk}\gamma_k\bigg{[}\frac{i}{[\frac{\Gamma_0}{2}+i(\Omega-\omega_k)](\omega_k-\omega_j)}(e^{i(\omega_j-\omega_k)t}-1)\nonumber\\&&+\frac{1}{[\frac{\Gamma_0}{2}+i(\Omega-\omega_k)][\frac{\Gamma_0}{2}+i(\Omega-\omega_j)]}(e^{i[\frac{\Gamma_0}{2}+i(\Omega-\omega_j)]t}-1)\bigg{]}.
	%\end{eqnarray}
	This equation corresponds to the lowest order approximation in the qubit-bath coupling. The population $|\mathcal{C}_j(t)|^2$ reads now
	\begin{eqnarray}\label{cjt-amp2}
		|\mathcal{C}_j(t)|^2&&\simeq \frac{ \gamma_j^2}{\hbar^2} \frac{(1-e^{-[\frac{\Gamma_0}{2}+i(\Omega-\omega_j)]t})(1-e^{-[\frac{\Gamma_0}{2}-i(\Omega-\omega_j)]t})}{\frac{\Gamma_0^2}{4}+(\Omega-\omega_j)^2}\nonumber\\&&=\frac{ \gamma_j^2}{\hbar^2}\,\frac{1-2e^{-\frac{\Gamma_0}{2}t}\cos[(\Omega-\omega_j)t]+e^{-\Gamma_0t}}{\frac{\Gamma_0^2}{4}+(\Omega-\omega_j)^2}.
	\end{eqnarray}
	At $t\rightarrow \infty$, we have
	\begin{eqnarray}\label{result_final}
		|\mathcal{C}_j(t\rightarrow \infty)|^2=\frac{4\gamma_j^2}{\hbar^2\Omega^2}\,\frac{1}{(\frac{\Gamma_0}{\Omega})^2+4(1-\frac{\omega_j}{\Omega})^2}.
	\end{eqnarray}
	%Based on presented expressions for $\Gamma_0=\frac{2\pi}{\hbar^2}\nu_0\langle \gamma_j^2\rangle$ (Eq.~\eqref{m25}) and $\nu_0=N/(2\Omega)$, the average population $\langle |\mathcal{C}_j(t)|^2\rangle$ at this limit yields
	%\begin{eqnarray}\label{ave-pop}
	%\langle |\mathcal{C}_j(t\rightarrow \infty)|^2\rangle=\frac{4}{N\pi}\,\frac{\Gamma_0/\Omega}{(\frac{\Gamma_0}{\Omega})^2+4(1-\frac{\omega_j}{\Omega})^2}.
	%\end{eqnarray}
	This expression preserves normalization, $\sum_{j=1}^N |\mathcal{C}_j(t\rightarrow \infty)|^2 =1$, and it yields the correct energy of the microcanonical system, $\sum_{j=1}^N |\mathcal{C}_j(t\rightarrow \infty)|^2 \hbar\omega_j =\hbar\Omega$, while the qubit energy approaches zero in the long time limit. For $\Gamma_0 \rightarrow 0$, the distribution becomes a delta function at $\omega_j=\Omega$. The expectation value of this quantity for the spin $j$ is then given by
	\begin{eqnarray} \label{result_final2}
		\langle|\mathcal{C}_j(t\rightarrow \infty)|^2\rangle=\frac{4}{N \pi}\,\frac{\frac{\Gamma_0}{\Omega}}{(\frac{\Gamma_0}{\Omega})^2+4(1-\frac{\omega_j}{\Omega})^2}.
	\end{eqnarray}
	
	Equation~\eqref{result_final} states that the populations of the bath spins stabilize at a fixed value given by a Lorentzian distribution at infinite time. This is a clear signature of irreversibility in the model we are considering, and it is in contrast with the fact that our model is integrable. Still, this result is physically meaningful, as we discuss in the following. First, let us  recall the assumptions we have made to obtain Eq.~\eqref{result_final}:
	\begin{enumerate}[i)]
		\item There are no direct interactions between the bath spins, i.e., $\kappa_{jk}=0$. This assumption can be easily relaxed, as the bath Hamiltonian in the presence of direct interactions can be diagonalized to obtain a new Hamiltonian describing a collection of non-interacting modes, as discussed in Appendix~\ref{appendix:integrability}.
		\item There is weak-coupling between the qubit and the spins, i.e., $\gamma_j$ are a perturbation of the model Hamiltonian.
		\item The values of the qubit-spin couplings and the spin energies are not correlated. This assumption has been taken only for the sake of simplicity. Even if this is not the case, the final conclusions do not change.
		\item The number of spins is large, i.e., $N\gg 1$. This is a key assumption we need in order to transform the summation over the different spins into an integral, which is the mathematical procedure that is giving rise to irreversibility in the population dynamics.
		\item The frequency range is broad enough so that there exists a timescale for which $\Delta \omega t \gg 1$. 
	\end{enumerate}
	The last two assumptions are crucial to understand how dissipation and irreversibility are emerging in the model we are considering. $N\gg 1$ secures that the spins have a ``dense enough'' distribution of frequencies. The limit $\Delta\omega t\gg 1$ is extending the integration interval for $\chi(t)$ in Eq.~\eqref{eqn:chiIntegral} from $-\infty$ to $\infty$. Therefore, these approximations correspond to treating the open system dynamics as if the qubit was interacting with an infinite collection of spins in the weak-coupling regime, which is exactly the standard Markovian model we have described in Sec.~\ref{sec:Ther-qubit}. Hence, it is no surprise that the qubit is decaying exponentially as in the Markovian case, the only difference being in the expression for the decay rate in Eq.~\eqref{rate}. Nonetheless, contrary to the Markovian treatment, in our approach we are also following the dynamics of each individual spin of the bath without ignoring the back-reaction of the qubit, and we can find an approximate solution for their long-time populations, given by Eq.~\eqref{result_final}. This long-time solution of the dynamics is clearly just an approximation, because we know that the model is integrable and there will exist a time at which the system goes back to its initial conditions. 
	
	\subsection{Numerical results and comparison with the analytical limit}

\begin{figure}[h!]
		\centering
		\includegraphics [width=0.8\columnwidth] {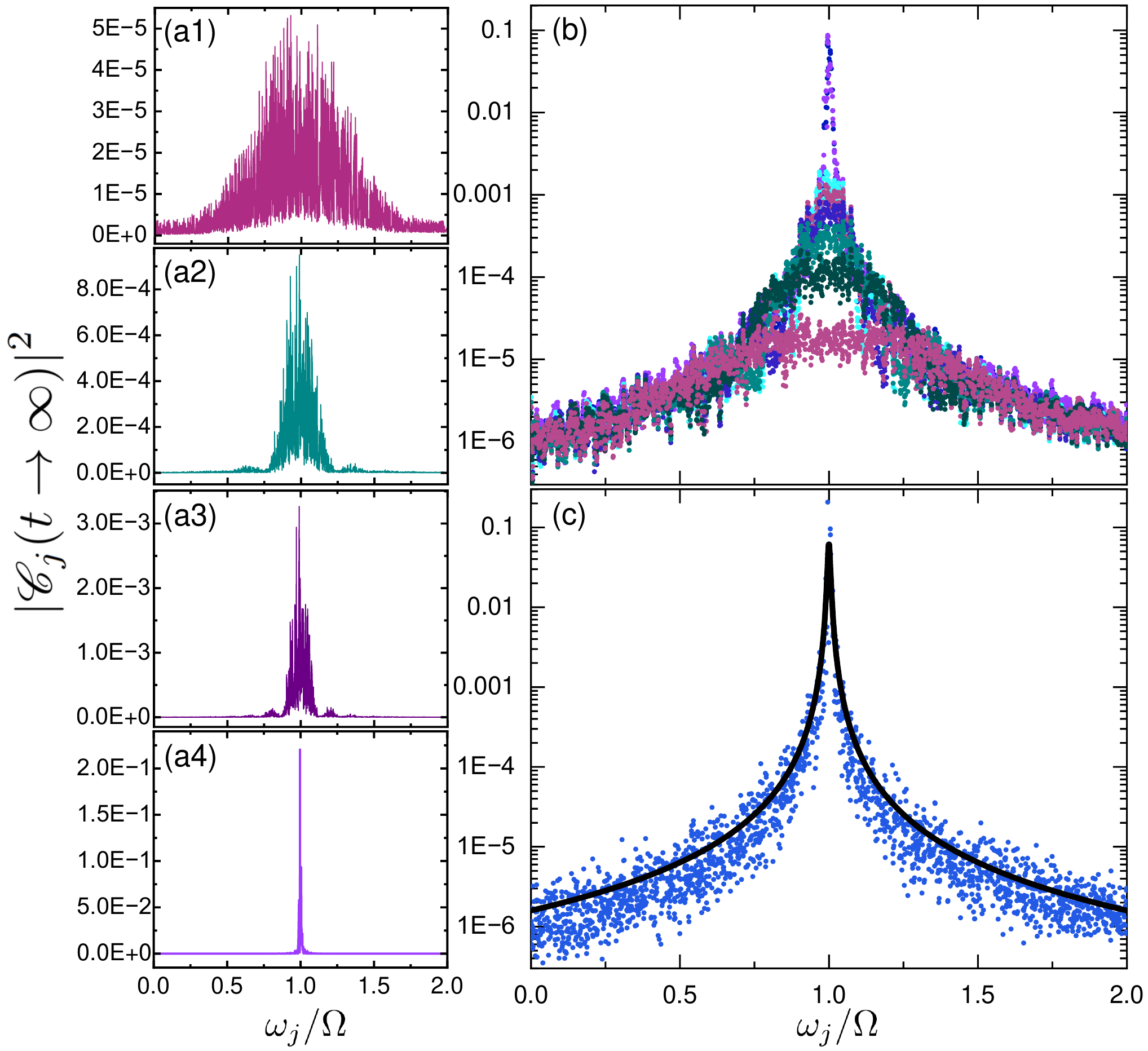}
		\caption{Results of relaxation to an equilibrium state when the bath is initially in the ground state and the qubit is in the excited state. (a) Snapshots of the distribution of the populations of the spins in four different short time intervals, $\Omega t < 100$, $200 < \Omega t < 300$, $400 < \Omega t < 500$, and $9900 < \Omega t < 10000$. (b) The same as (a) but all in the same panel and on the logarithmic scale with a more complete set of eight time intervals. (c) The long time populations in the interval $9500 < \Omega t < 10000$. The symbols show the exact numerical result for the bath spins, and the solid line is the analytic approximation of Eq.~\eqref{result_final2} for the asymptotic long time limit. We have assumed $N=2000$, $\Gamma_0/\Omega=0.01$, and uniform and uncorrelated distributions for $\gamma_j$:s and $\omega_j$:s. 		\label{Fig-population}}
	\end{figure}

	For the numerical result, we integrate Eq.~\eqref{cit} in time $t$ yielding the exact solution of the Schr\"odinger equation for our model. Some results of these numerical calculations are summarized in Fig. \ref{Fig-population}. For these data, we have chosen $\Gamma_0/\Omega = 0.01$, $N=2000$, and the distributions of $\gamma_j$ and $\omega_j$ are uniform and uncorrelated as described above, with $\Delta \omega /2 = \Omega$. Furthermore, we have included the mutual couplings $\kappa_{jk}$ with the same distribution as $\gamma_j$. The time integral is extended from 0 to $\Omega t = 10^4$. The results are presented in form of individual spin populations $|\mathcal{C}_j(t)|^2$ for all the spins as functions of their corresponding energy $\hbar\omega_j$. In the figure panels the energies have been normalized by that of the qubit, i.e. the horizontal axes are shown in form of $\omega_j/\Omega$. 
	
	In Fig. \ref{Fig-population}(a) we present the time evolution of the spin populations; here the vertical $|\mathcal{C}_j(t)|^2$ scale is linear. The populations are averaged over short time intervals demonstrating how initially at short times, $t \ll \Gamma_0^{-1}$, all the spins are effectively in the ground state, whereafter they receive the energy of the qubit and get excited. Remarkably the distribution is initially quite broad in $\omega_j$ but sharpens towards the asymptotic Lorentzian distribution when $t > \Gamma_0^{-1}$. The main frame is Fig. \ref{Fig-population}(b) shows a more complete set (eight snapshots) as in (a) but now on logarithmic scale and all time intervals combined. In (c) we demonstrate the main result by contrasting the exact numerical populations in the long time limit, averaged over $9500 < \Omega t \le 10000$ against the analytic Lorentzian distribution (solid line from Eq. \eqref{result_final2}). 
	
	The analytical approximation of Eq. \eqref{result_final2} compares remarkably well with the result of long-time numerical simulations in Fig. \ref{Fig-population}(c). In practice this means that we can essentially reach the dissipative large $N$ limit with a limited number of spins. 
	
	\section{Final remarks} 
	\label{sec:conclusions}

	The emergence of effective relaxation in the physical model we have studied is due to the fact that we have considered the long-time dynamics of some quantities, namely the populations of each two-level system, that are functions of many ($N\gg 1$) different oscillating normal modes of the system, as explained in Appendix A. This concept is related to Khinchin's approach to thermalization in closed classical systems \cite{khinchin}, according to which thermalization can occur also in a perfectly integrable model, if we only look at specific observables that are ``sum functions'', i.e., sum of many decoupled canonical variables of the integrable system. In the model we have studied, however, we observe relaxation without thermalization. For future studies, it may be interesting to investigate the relation between our results and Khinchin's thermalization, which has been recently observed numerically in classical models that are quite similar to the dynamics in the single-excitation sector we have analysed in this work \cite{2021JSP...183...41B,COCCIAGLIA2022127581}.
	
	Obviously our model does not demonstrate thermalization into Gibbs distribution. Quantum thermalization \cite{Gogolin2016,Popescu2006,Lebowitz2006,dalessio2016,deutsch2018,Mori_2018,Nandkishore2015,Reimann2016,Chen2021} often counts on ergodicity and nearly exponential increase of available microstates with increasing energy. In the example of zero-temperature initial state that we have discussed we only probe states with single-excitation because of the number-conserving form of the coupling Hamiltonian. This conserved number of excitations does not comply with the exponential increase of number of states with energy. It fails to satisfy the ergodicity requirement of thermalization. Secondly, we did not expand, analytically, beyond the golden rule treatment. So we anticipate that in order to see full thermalization, one needs to consider the next order process (two photons) to couple spins with different energies and a coupling Hamiltonian where double creation and annihilation operations are possible, like $(a+a^\dagger)(b_i+b_i^\dagger)$ and $(b_i+b_i^\dagger)(b_j+b_j^\dagger)$. This interaction, unlike the one in rotating wave approximation, modifies the ground state of the model as briefly discussed in Appendix~\ref{appendix:groundstate}. Moreover, this form arises naturally for instance in circuit QED setups for superconducting qubits with inductive and capacitive coupling. However, one needs an alternative method to address this regime due to the exponential expansion of the Hilbert space.

	\section{Acknowledgments}
	We thank Paolo Muratore-Ginanneschi, Ivan Khaymovich, Charles Marcus, Joachim Ankerhold, and Erik Aurell for useful discussions. This work was funded through Academy of
	Finland grant 312057.
	
	\appendix
	
	\section{Proof of integrability of the model in the single-excitation sector}
	\label{appendix:integrability}
	In the single-excitation sector spins and bosonic harmonic oscillators are indistinguishable. Therefore, for convenience, in this appendix we work with the spin operators $a$ and $b_j$ as if they were bosonic operators. Then, we can write the model Hamiltonian in the Schr\"{o}dinger picture introduced in Sec.~\ref{sec:model} as a quadratic Hamiltonian represented by
	\begin{equation}
		\label{eqn:quadraticHam}
		H = \mathbf{v}^T \mathsf{M_H} \,\mathbf{w},
	\end{equation}
	where $\mathbf{v}=(a^\dagger,b_1^\dagger,b_2^\dagger,\ldots,b_N^\dagger)^T$ is the vector of bosonic creation operators, and equivalently $\mathbf{w}=(a,b_1,b_2,\ldots,b_N)^T$. In this representation, the Hamiltonian matrix is given by (we remind that $\gamma_j$ and $\kappa_{jk}$ are real):
	\begin{equation}
		\label{eqn:matrixHam}
		\mathsf{M_H} = \begin{pmatrix}
			\Omega & \gamma_1 & \gamma_2& \ldots & \ldots & \gamma_N\\
			\gamma_1 & \omega_1 & \kappa_{12}& \kappa_{13} & \ldots & \kappa_{1N}\\
			\gamma_2 & \kappa_{12} & \omega_2 & \kappa_{23} & \ldots & \kappa_{2N}\\
			\gamma_3 & \kappa_{13} & \kappa_{23} & \omega_3 & \ldots & \vdots\\
			\vdots & \vdots & \vdots & \vdots & \ddots & \vdots\\
			\gamma_N & \kappa_{1N} & \kappa_{2N} & \ldots & \ldots & \omega_N\\
		\end{pmatrix}.
	\end{equation}
	This matrix can also be taken as the Hamiltonian of the model in the single-excitation sector, that is, in the basis $\{\ket{j}\}_{j=0}^N$ we have introduced in Sec.~\ref{sec:singleEx}.
	
	$\mathsf{M_H}^\dagger=\mathsf{M_H}$, therefore it can be diagonalized by a unitary transformation $U$ as $\mathsf{\tilde{M}_H}=U \mathsf{M_H} U^\dagger =\text{diag}(f_0,\ldots,f_{N})$, while the new vectors of bosonic creation and annihilation operators are given by $\tilde{\mathbf{w}}=U\mathbf{w}$, $\tilde{\mathbf{v}}=U^*\mathbf{v}$ ($U^*$ is the complex conjugate of $U$). Indeed, it is easily verified that $[\tilde{v}_j,\tilde{w}_k]=\sum_{m,n=0}^{N}U_{kn}U^\dagger_{mj}[v_m,w_n]=\delta_{jk}$. Then, the Hamiltonian is written as
	$H=\tilde{\mathbf{v}}^T \mathsf{\tilde{M}_H} \,\tilde{\mathbf{w}}$, and it describes a collection of $N+1$ bosonic modes. Suppose that $\mathbf{\tilde{w}}=(c_0,c_1,\ldots,c_{N})$, and equivalently for $\mathbf{\tilde{v}}$. Then,
	\begin{equation}
		H = \sum_{j=0}^{N} H_j =\sum_{j=0}^{N} f_j c_j^\dagger c_j,
	\end{equation}
	where the eigenvalues of the matrix $\mathsf{M_H}$ are the frequencies of the decoupled bosonic modes $c_j$. Each single-mode Hamiltonian $H_j$ is therefore a conserved quantity of the dynamics. Since there are $N+1$ commuting and independent conserved quantities the model is trivially integrable, as the solution of the dynamics at time $t$ is just given by the proper linear combination of modes $c_j$ rotating with frequency $f_j$. 
	
	Finally, note that we may have also diagonalized only the matrix of the bosonic modes of the bath in Eq.~\eqref{eqn:matrixHam} (i.e., the square matrix obtained by removing the first row and column). In this sense, the inter-spin couplings $\kappa_{jk}$ are only modifying the form of the frequencies $\omega_j$ and of the qubit-bath couplings $\gamma_j$, but do not change the structure of the physical model we are considering.
	
	\section{Does our model comply with the ground state of the spin system being that of each spin individually in the ground state?}
	\label{appendix:groundstate}
	If we consider a system composed of many isolated spins, the ground state by default will be $|0\,0\,0\,...\,0\rangle$, meaning that all the spins are in their ground state. But if we consider couplings between the spins, there will be a superposition between the spins due to the coupling (although the contributions are very small in the weak coupling limit), and the eigenstates are of the form $\eta|0\,0\,0\,...\,0\rangle+\alpha|1\,0\,1\,...\,0\rangle+\beta|0\,1\,...1\,0\rangle+\gamma|1\,0\,...0\,0\rangle...$ where $\alpha,~\beta,~\gamma,~...\ll 1$, and $\eta~\approx 1$. In order to test this idea, we simply assume the two-spin model, where the perturbation is given by $\mathbb{\hat{V}}_{\rm JC}=g(\hat{a}_1^\dagger\hat{a}_2+\hat{a}_1\hat{a}_2^\dagger)$. In this case the Hamiltonian in the basis $\{|00\rangle,|10\rangle,|01\rangle,|11\rangle\}$ is given by 
	\begin{eqnarray}\label{JC.2spins}
		\left[\begin{array}{cccc}
			0 & 0 & 0 & 0	\\
			0 & \hbar\omega_1 & g & 0	\\
			0 & g & \hbar\omega_2 & 0	\\
			0 & 0 & 0 & \hbar\omega_1+\hbar\omega_2
		\end{array}\right],
	\end{eqnarray}
	meaning that even in the presence of the coupling still the ground state is $|0\,0\,0\,...\,0\rangle$. This is true for any Hamiltonian in the rotating wave approximation, as the annihilation operators in the interaction term are always yielding zero on the state $|0\,0\,0\,...\,0\rangle$, and thus they do not modify the ground state.
	
	If we instead consider a more complete form of the perturbation, $\mathbb{\hat{V}}=g(\hat{a}_1+\hat{a}_1^\dagger)(\hat{a}_2+\hat{a}_2^\dagger)$, the Hamiltonian in the same basis as above is given by
	\begin{eqnarray}\label{H.2spins}
		\left[\begin{array}{cccc}
			0 & 0 & 0 & g	\\
			0 & \hbar\omega_1 & g & 0	\\
			0 & g & \hbar\omega_2 & 0	\\
			g & 0 & 0 & \hbar\omega_1+\hbar\omega_2
		\end{array}\right],
	\end{eqnarray}
	which then gives us hybridized eigenstates as expected.

\end{document}